\documentclass{moriond}

\def\be{\begin{equation}}
\def\ee{\end{equation}}
\def\bea{\begin{eqnarray}}
\def\eea{\end{eqnarray}}

\newcommand{\tsys}{T_\mathrm{sys}}

\newcommand{\Photo}{\includegraphics[height=35mm]{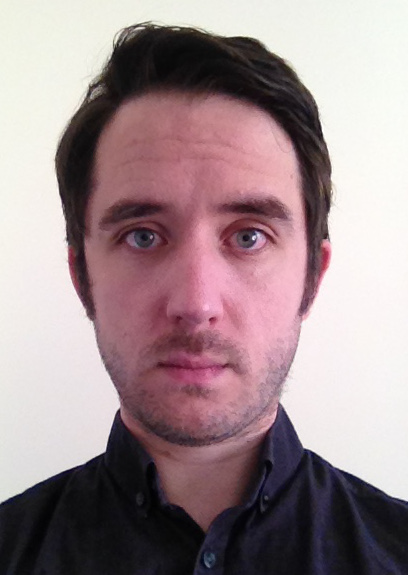}}

\begin{document}
\vspace*{4cm}
\title{Simulation of Systematics in Future Single-Dish HI Intensity Mapping Experiments}

\author{S. E. Harper on behalf of the Manchester Intensity Mapping Team}

\address{Jodrell Bank Centre for Astrophysics, School of Physics and Astronomy, Alan Turing Building, University of Manchester, Manchester, M13 9PL UK}

\maketitle\abstracts{
HI intensity mapping (IM) is an exciting new probe that could revolutionize the future of cosmology. However, the relative faintness of the HI signal when compared to foregrounds of astrophysical or terrestrial origin will make HI IM extremely challenging. The imprint of these foregrounds may result in systematic errors in the recovered cosmological signal. We discuss an IM simulation pipeline developed at Manchester that can introduce systematic errors at the TOD level in order to help assess their impact. We will present results for two potential sources of systematics for HI IM surveys: $1/f$ noise and the integrated emission from global navigation satellites.
}

\section{Introduction}\label{sec:intro}

HI intensity mapping (IM) is a promising new cosmological probe (see\cite{Battye2004,Battye2013,Bull2015} and references therein) by having the potential to map fluctuations in the cosmic matter density across a huge span of cosmic time\cite{Loeb2004,Pritchard2008,Breysse2018}.

The expected scale of the HI IM fluctuations are of the order 100\,$\mu$K\cite{Battye2004}, which is several orders of magnitude fainter than the known brightness of astrophysical foregrounds. The huge difference in brightness between the HI signal and foregrounds puts extreme constraints on instrumentation to be both spectrally and temporally stable, as most current techniques for foreground subtraction rely upon the spectral smoothness of astrophysical foregrounds\cite{Liu2011,Alonso2015,BigotSazy2015}. 

There are many possible systematic within a HI IM dataset that may result in the foregrounds signals deviating from spectral smoothness. Many of these systematics occur due to the instrumentation (such as $1/f$ gain fluctuations\cite{BigotSazy2015,Harper2017}, beam sidelobes or polarisation leakage\cite{Alonso2015}), or the environment (e.g., man-made radio-frequency interference or RFI). Many of these systematics are temporally variable, and therefore assessing the impact of them for future HI IM surveys requires end-to-end simulations of observations at the time-ordered data (TOD) level. In these proceedings we will present just such a pipeline and the current status of assessing the impact of $1/f$ noise and RFI from  global navigation satellites on future single-dish HI IM surveys.

\section{Simulations}\label{sec:simulations}

We have developed a simulation pipeline for modelling a general single-dish HI IM experiment at the level of the generation of the TOD. The pipeline includes methods for simulating the expected cosmological HI signal, maps of Galactic foregrounds, and a suite of experiment specific systematics that act at the TOD level. The pipeline also includes methods for data analysis and processing, and includes both basic component separation methods such as PCA\cite{BigotSazy2015,Alonso2015} and advanced novel methods such as GNILC\cite{Remazeilles2011,Olivari2016}. The pipeline is written in a combination of \texttt{Python}, \texttt{C} and \texttt{FORTRAN}, and is designed to be fully parallelised using \texttt{MPI} libraries. 

\subsection{1/f Noise}

$1/f$ noise is a familiar systematic for single-dish radio and sub-mm astronomy\cite{Seiffert2002,Cantalupo2010}. For HI IM experiments $1/f$ noise originating from gain fluctuations in receiver amplifiers should be very spectrally correlated, and as such should be easily removed using existing foreground subtraction methods\cite{Harper2017}. However, small deviations in the spectral smoothness of the $1/f$ noise, introduced by either the instrumentation or data analysis methods, greatly increases the difficulty in separating the $1/f$ noise from the HI signal. 

\begin{figure}
    \centering
    \includegraphics[width=\textwidth]{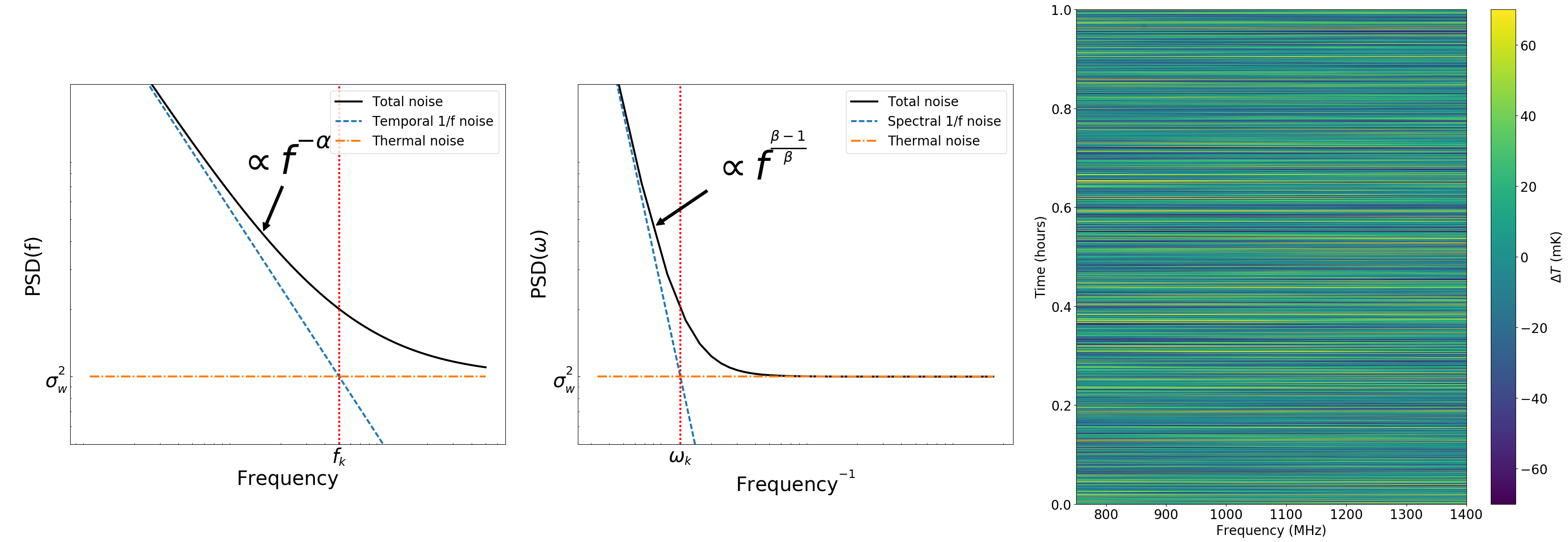}
    \caption{Model power spectra of $1/f$ noise that has correlated fluctuations in both time and frequency (\textit{left} and \textit{middle}). The $1/f$ noise correlations in time are described by the parameter $\alpha$, and in frequency by $\beta$. The knee frequency $f_k$ (or in frequency $\omega_k$), is where the power of the $1/f$ noise spectrum and white noise power ($\sigma_w^2$) are equal. An example waterfall plot of $1/f$ noise from the IM simulation pipeline is shown in the \textit{right} plot for $\alpha = 1$, $\beta =0.25$ and a system temperature of $\tsys = 20$\,K. }
    \label{fig:fig0}
\end{figure}

Fig.~\ref{fig:fig0} shows the model power spectrum of $1/f$ noise in both time and frequency. The correlations in time are described by the index $\alpha$ and in frequency by $\beta$, the amplitude of the fluctuations is dictated by the knee frequency $f_k$ (the temporal scale where the power density of the $1/f$ noise and white noise are unity). The waterfall plot shows an example output from the simulations for a $\beta = 0.25$. In a recent paper\cite{Harper2017} various values of $\beta$ between 0 (completely correlated) to 1 (completely uncorrelated) were explored in the context of a model phase 1 SKA-MID array. It was found that for even very small levels of \textit{decorrelation} in the $1/f$ noise spectrum (e.g., $\beta > 0.25$) results in a greatly increased uncertainty in the recovered HI power spectrum on large scales. If the $1/f$ noise is entirely uncorrelated with a knee frequency close to $1\,Hz$ per 20\,MHz channel width it was found that a 30\,day SKA HI IM survey would not even be able to detect the cosmological HI signal on any scale.

\subsection{Global Navigation Satellites}

\begin{figure}
    \centering
    \includegraphics[width=\textwidth]{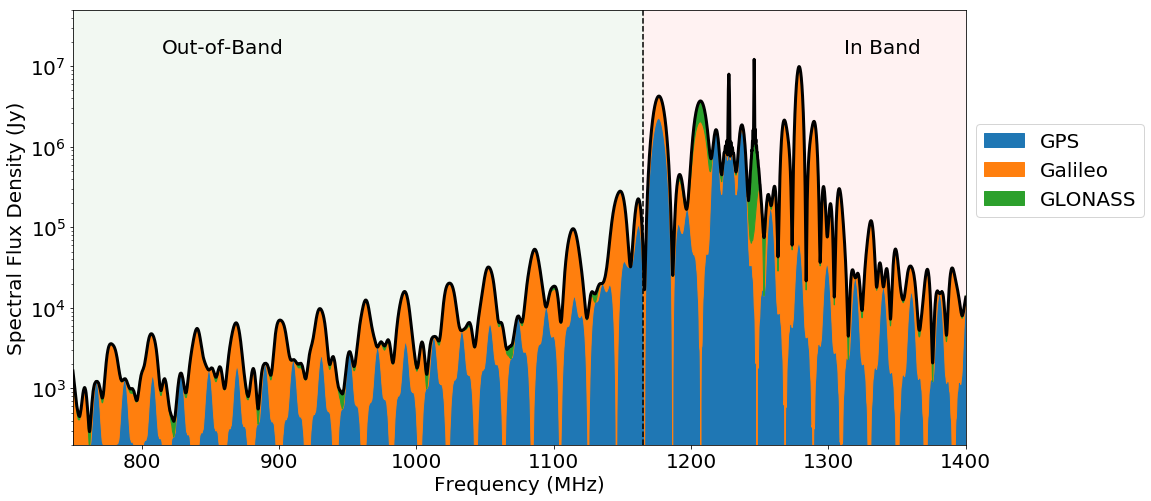}
    \caption{Expected flux density contributions of of GNSS satellites for the GPS, Galileo and GLONASS constellations if all three satellites were within the main beam of the telescope. Region marked \textit{in band} is the official GNSS allocation, while the \textit{out-of-band} region is the predicted leakage of GNSS power into non-GNSS allocations. For reference the radio flux density of the quiet Sun is between $10^5 - 10^6$\,Jy.}
    \label{fig:fig1}
\end{figure}

The global navigation satellites system (GNSS) is a network of approximately 120 satellites spread predominantly over the three global constellations: GPS (USA), GLONASS (Russia) and Galileo (Europe). Alongside these there are also more regional constellations controlled by Japan, India and China. Currently there are approximately 60 GNSS satellites in orbit with plans to expand to 120 by 2030\cite{Gao2012}. Each one of these satellites usually has three services that broadcast with central frequencies between $1165 < \nu < 1610$\,MHz\cite{hofmann2007}, however the nature of the GNSS transmission means that power leaks into out-of-bands regions of the spectrum that, although are well within international limits\cite{ITU2015}, are still bright enough to be problematic for HI radio astronomy. This is because each one of these satellites are as bright as the quiet Sun when observed within the GNSS band\cite{steigenberger2017,Harper2018}, but unlike the Sun there are always 6 or more satellites above the horizon at any given time all moving continuously within the sidelobes of the telescope. Fig.~\ref{fig:fig1} shows an example of the expected GNSS spectrum when observing directly one satellite from each GNSS constellation and highlights the complex spectral structure of the GNSS satellite emission.

As the GNSS satellites are moving in fixed orbital planes that are comoving with the celestial sky the integrated emission of the satellites over a long HI IM survey would stack. The impact of this integrated emission was assessed using the IM pipeline and assuming a model HI IM survey with the phase 1 SKA-MID array\cite{Harper2018}. The nature of the integrated GNSS emission is a convolution of the satellite celestial tracks and the beam (specifically the far sidelobes) of the observing telescope. It was shown recently\cite{Harper2018} that an SKA HI IM survey will have a substantial contribution from the integrated GNSS emission at frequencies $\nu > 900$\,MHz or redshifts of $z < 0.6$. This is concerning as the integrated GNSS emission \textit{foreground} will not be spectrally smooth (as shown in Fig.~\ref{fig:fig1}), therefore existing component separation methods may find it challenging to remove and a new, novel approach may be required.

\section{Conclusion}

HI IM is an upcoming new method with great potential as a cosmological probe. However the faintness of the HI signal relative to astrophysical, instrumental and man-made \textit{foregrounds} will present a serious challenge for future experiments and surveys. In these proceedings we have discussed the Manchester IM simulations pipeline and how it will be useful for assessing the impact of systematic errors that act at the TOD level, such as $1/f$ noise, beam sidelobes, or polarisation leakage. We have specifically discussed the impact of $1/f$ noise, which in the best case scenario may be far less of a problem than in previous CMB experiments but requires careful processing of the data and instrumental design. We also discussed the impact of the integrated emission from GNSS satellites, a potentially unique foreground to HI IM experiments, and how it will require a novel approach to remove from observations at redshifts $z < 0.6$.

\section*{Acknowledgments}

SH acknowledges support from an STFC Consolidated Grant (ST/P000649/1). SH also acknowledges support from an ERC Starting (Consolidator) Grant (no.$\sim$307209) under FP7. 

\section*{References}
\footnotesize
\bibliography{references}

\begin{thebibliography}{10}

\bibitem{Battye2004}
R.~A. {Battye}, R.~D. {Davies}, and J.~{Weller}.
\newblock {Neutral hydrogen surveys for high-redshift galaxy clusters and
  protoclusters}.
\newblock {\em \mnras}, 355:1339--1347, December 2004.

\bibitem{Battye2013}
R.~A. {Battye}, I.~W.~A. {Browne}, C.~{Dickinson}, G.~{Heron}, B.~{Maffei}, and
  A.~{Pourtsidou}.
\newblock {H I intensity mapping: a single dish approach}.
\newblock {\em \mnras}, 434:1239--1256, September 2013.

\bibitem{Bull2015}
P.~{Bull}, P.~G. {Ferreira}, P.~{Patel}, and M.~G. {Santos}.
\newblock {Late-time Cosmology with 21 cm Intensity Mapping Experiments}.
\newblock {\em \apj}, 803:21, April 2015.

\bibitem{Loeb2004}
A.~{Loeb} and M.~{Zaldarriaga}.
\newblock {Measuring the Small-Scale Power Spectrum of Cosmic Density
  Fluctuations through 21cm Tomography Prior to the Epoch of Structure
  Formation}.
\newblock {\em Physical Review Letters}, 92(21):211301, May 2004.

\bibitem{Pritchard2008}
J.~R. {Pritchard} and A.~{Loeb}.
\newblock {Evolution of the 21cm signal throughout cosmic history}.
\newblock {\em \prd}, 78(10):103511, November 2008.

\bibitem{Breysse2018}
P.~C. {Breysse}, Y.~{Ali-Ha{\"i}moud}, and C.~M. {Hirata}.
\newblock {The ultimate frontier of 21cm cosmology}.
\newblock {\em ArXiv e-prints}, April 2018.

\bibitem{Liu2011}
A.~{Liu} and M.~{Tegmark}.
\newblock {A method for 21 cm power spectrum estimation in the presence of
  foregrounds}.
\newblock {\em \prd}, 83(10):103006, May 2011.

\bibitem{Alonso2015}
D.~{Alonso}, P.~{Bull}, P.~G. {Ferreira}, and M.~G. {Santos}.
\newblock {Blind foreground subtraction for intensity mapping experiments}.
\newblock {\em \mnras}, 447:400--416, February 2015.

\bibitem{BigotSazy2015}
M.-A. {Bigot-Sazy}, C.~{Dickinson}, R.~A. {Battye}, I.~W.~A. {Browne}, Y.-Z.
  {Ma}, B.~{Maffei}, F.~{Noviello}, M.~{Remazeilles}, and P.~N. {Wilkinson}.
\newblock {Simulations for single-dish intensity mapping experiments}.
\newblock {\em \mnras}, 454:3240--3253, December 2015.

\bibitem{Harper2017}
S.~{Harper}, C.~{Dickinson}, R.~A. {Battye}, S.~{Roychowdhury}, I.~W.~A.
  {Browne}, Y.-Z. {Ma}, L.~C. {Olivari}, and T.~{Chen}.
\newblock {Impact of Simulated 1/f Noise for HI Intensity Mapping Experiments}.
\newblock {\em \mnras}, May 2018.

\bibitem{Remazeilles2011}
M.~{Remazeilles}, J.~{Delabrouille}, and J.-F. {Cardoso}.
\newblock {Foreground component separation with generalized Internal Linear
  Combination}.
\newblock {\em \mnras}, 418:467--476, November 2011.

\bibitem{Olivari2016}
L.~C. {Olivari}, M.~{Remazeilles}, and C.~{Dickinson}.
\newblock {Extracting H I cosmological signal with generalized needlet internal
  linear combination}.
\newblock {\em \mnras}, 456:2749--2765, March 2016.

\bibitem{Seiffert2002}
M.~{Seiffert}, A.~{Mennella}, C.~{Burigana}, N.~{Mandolesi}, M.~{Bersanelli},
  P.~{Meinhold}, and P.~{Lubin}.
\newblock {1/f noise and other systematic effects in the Planck-LFI
  radiometers}.
\newblock {\em \aap}, 391:1185--1197, September 2002.

\bibitem{Cantalupo2010}
C.~M. {Cantalupo}, J.~D. {Borrill}, A.~H. {Jaffe}, T.~S. {Kisner}, and
  R.~{Stompor}.
\newblock {MADmap: A Massively Parallel Maximum Likelihood Cosmic Microwave
  Background Map-maker}.
\newblock {\em \apjs}, 187:212--227, March 2010.

\bibitem{Gao2012}
Grace~Xingxin Gao and Per Enge.
\newblock How many gnss satellites are too many?
\newblock {\em IEEE Transactions on aerospace and electronic Systems},
  48(4):2865--2874, 2012.

\bibitem{hofmann2007}
Bernhard Hofmann-Wellenhof, Herbert Lichtenegger, and Elmar Wasle.
\newblock {\em GNSS--global navigation satellite systems: GPS, GLONASS,
  Galileo, and more}.
\newblock Springer Science \& Business Media, 2007.

\bibitem{ITU2015}
International~Telecommunication Union.
\newblock Unwanted emissions in the out-of-band domain., 2015.

\bibitem{steigenberger2017}
Peter Steigenberger, Steffen Thoelert, and Oliver Montenbruck.
\newblock Gnss satellite transmit power and its impact on orbit determination.
\newblock {\em Journal of Geodesy}, pages 1--16, 2017.

\bibitem{Harper2018}
S.~{Harper} and C.~{Dickinson}.
\newblock {Potential Impact of Global Navigation Satellite Services on Total
  Power HI Intensity Mapping Surveys}.
\newblock {\em ArXiv e-prints}, March 2018.

\end{thebibliography}

\end{document}